\documentclass{emulateapj}
\usepackage{apjfonts}
\usepackage{comment}
\usepackage{units}
\usepackage{gensymb}
\usepackage{amsmath}
\usepackage{subfigure}
\usepackage{graphicx}
\usepackage{paralist}

\newcommand{\Msun}{$\rm M_{\odot}$}
\newcommand{\Lsun}{$\rm L_{\odot}$}
\newcommand{\us}{$\mu$s}
\newcommand{\dmu}{pc~cm$^{-3}$}
\newcommand{\psr}{PSR J1930$-$1852}
\newcommand{\jname}{J1930$-$1852}

\begin{document}
\title{\psr: a pulsar in the widest known orbit around another neutron star}
\author{
J.~K.~Swiggum\altaffilmark{1}, 
R.~Rosen\altaffilmark{1,2}, 
M.~A.~McLaughlin\altaffilmark{1}, 
D.~R.~Lorimer\altaffilmark{1}, 
S.~Heatherly\altaffilmark{2}, 
R.~Lynch\altaffilmark{3}, 
S.~Scoles\altaffilmark{2},
T.~Hockett\altaffilmark{4,5},
E.~Filik\altaffilmark{4},
J.~A.~Marlowe\altaffilmark{4},
B.~N.~Barlow\altaffilmark{4},
M.~Weaver\altaffilmark{6},
M.~Hilzendeger\altaffilmark{7}, 
S.~Ernst\altaffilmark{8}, 
R.~Crowley\altaffilmark{7}, 
E.~Stone\altaffilmark{8}, 
B.~Miller\altaffilmark{9}, 
R.~Nunez\altaffilmark{10}, 
G.~Trevino\altaffilmark{6}, 
M.~Doehler\altaffilmark{9}, 
A.~Cramer\altaffilmark{11}, 
D.~Yencsik\altaffilmark{7}, 
J.~Thorley\altaffilmark{1,9}, 
R.~Andrews\altaffilmark{9}, 
A.~Laws\altaffilmark{12}, 
K.~Wenger\altaffilmark{6},
L.~Teter\altaffilmark{6}, 
T.~Snyder\altaffilmark{10}, 
A.~Dittmann\altaffilmark{13}, 
S.~Gray\altaffilmark{10}, 
M.~Carter\altaffilmark{9}, 
C.~McGough\altaffilmark{9}, 
S.~Dydiw\altaffilmark{7},
C.~Pruett\altaffilmark{8}, 
J.~Fink\altaffilmark{6},
A.~Vanderhout\altaffilmark{14}
}

\altaffiltext{1}{Dept.~of Physics \& Astronomy, WVU, Morgantown, WV 26506, USA}
\altaffiltext{2}{NRAO, P.O. Box 2, Green Bank, WV 24944, USA}
\altaffiltext{3}{Dept. of Physics, McGill Univ., Montreal, QC H3A 2T8, Canada}
\altaffiltext{4}{Dept.~of Physics, High Point Univ., High Point, NC 27268, USA}
\altaffiltext{5}{Dept.~of Physics and Optical Science, UNC at Charlotte, Charlotte, NC 28223, USA}
\altaffiltext{6}{Broadway H. S., 269 Gobbler Dr., Broadway, VA 22815, USA}
\altaffiltext{7}{Trinity H. S., 231 Park Ave., Washington, PA 15301, USA}
\altaffiltext{8}{Rowan County H. S., 499 Viking Dr., Morehead, KY 40351, USA}
\altaffiltext{9}{Strasburg H. S., 250 Ram Drive, Strasburg, VA 22657, USA}
\altaffiltext{10}{Hedgesville H. S., 109 Ridge Rd. N., Hedgesville, WV 25427, USA}
\altaffiltext{11}{Langley H. S., 6520 Georgetown Pike, McLean, VA 22101, USA}
\altaffiltext{12}{Central H. S., 1147 Susan Avenue, Woodstock, VA 22664, USA}
\altaffiltext{13}{George C. Marshall H. S., 7731 Leesburg Pike, Falls Church, VA 22043, USA}
\altaffiltext{14}{Nicolet H. S., 6701 N. Jean Nicolet Rd., Glendale, WI 53271, USA}

\begin{abstract}
In the summer of 2012, during a Pulsar Search Collaboratory workshop, two high-school students discovered \jname, a pulsar in a double neutron star (DNS) system. Most DNS systems are characterized by short orbital periods, rapid spin periods and eccentric orbits. However, \jname\ has the longest spin period (\unit[$P_{\rm spin}\sim$185]{ms}) and orbital period (\unit[$P_{\rm b}\sim$45]{days}) yet measured among known, recycled pulsars in DNS systems, implying a shorter than average and/or inefficient recycling period before its companion went supernova. We measure the relativistic advance of periastron for \jname, \unit[$\dot{\omega}=0.00078$(4)]{deg/yr}, which implies a total mass (\unit[M$_{\rm{tot}}=2.59$(4)]{\Msun}) consistent with other DNS systems. The $2\sigma$ constraints on M$_{\rm{tot}}$ place limits on the pulsar and companion masses (\unit[$m_{\rm p}<1.32$]{\Msun} and \unit[$m_{\rm c}>1.30$]{\Msun} respectively). \jname's spin and orbital parameters challenge current DNS population models and make \jname\ an important system for further investigation.
\end{abstract}

\section{Introduction}
To date, $\sim$2,300 pulsars are known \citep{psrcat} and $\sim$10\% of them are in binary systems, orbiting white dwarf (WD), neutron star (NS) or main sequence star (MS) companions. The vast majority of these binaries are NS-WD systems; many of these systems emerge from scenarios where the pulsar forms first, followed by its companion, which overflows its Roche Lobe; accretion transfers angular momentum to the pulsar, decreasing the spin period and resulting in a millisecond pulsar (MSP) orbiting a WD \citep{alp+82}. This process of accretion and spin-up is commonly referred to as {\it recycling} and the period derivative of a recycled pulsar tends to be significantly lower than that of an unrecycled pulsar with the same spin period. There are four observed examples of pulsars orbiting stars that have yet to evolve off the main sequence \citep{jml+92,kjb+94,sml+01,lyn+05}; an additional four have been found with planet-sized companions \citep{tat+93,wol+94,bbb+11,slr+14}. More massive companions end their evolution off the main sequence in supernovae, resulting in double neutron star (DNS) systems. DNS systems are far less likely to remain bound than NS-WD systems, since the former must survive two supernova explosions during formation. Only about 10\% of these binary systems remain bound after one supernova explosion \citep{bai+89}. The probability of remaining bound after two supernovae is much lower ($\sim$1\%) and only nine such systems have been found and studied previously (see references in Table \ref{tab:dns}).

DNS systems have tantalizing applications --- for example, testing theories of gravity by measuring relativistic effects \citep{fst+14,wnt+10,ksm+06} and predicting DNS merger rates relevant to ground-based gravitational wave detectors like LIGO \citep{kkl+10,kpm+14}. DNS systems have also provided some of the most precise NS mass measurements, allowing for a statistical investigation of the underlying mass distribution \citep{opn+12,spr+10,tc+99}.

\cite{tc+99} used a sample of 26 NSs (21 MSPs and five binary companion NSs) with measured masses to determine a mean NS mass, \unit[$\langle m\rangle=1.35\pm0.04$]{\Msun}. More recently, \cite{spr+10} argue that the underlying NS mass distribution for objects in DNS systems is bimodal, with narrow peaks at \unit[1.246]{\Msun} and \unit[1.345]{\Msun}. They also suggest that these peaks indicate unique formation scenarios, where the lower mass component represents NSs that formed via electron capture \citep{nom+84,plp+04} and the higher mass component is indicative of iron core-collapse \citep{ww+86}. \cite{opn+12} use a Bayesian statistical approach to infer mass distributions for NSs with distinct evolutionary histories; they agree that NS masses provide clues about respective formation scenarios, however, they express skepticism that NS mass distributions are as narrow as \cite{spr+10} claim. Therefore, additional information is necessary in identifying a NS's evolutionary history.

\cite{wwk+10} investigate core-collapse mechanisms in eight Galactic DNS systems by inferring progenitor mass of the second-born NS and the magnitude of the supernova kick it received at birth from measured DNS orbital parameters and kinematic information. Using these methods, they conclude that NS companions of PSRs B1534+12 and B1913+16 underwent iron core-collapse supernovae, while J0737$-$3039A's companion likely formed via electron capture supernova. This final result was corroborated by \cite{fsk+13}; through detailed pulse profile shape analysis, they constrained the double pulsar system geometry, concluding that the secondary supernova explosion was relatively symmetric, indicative of an electron capture process.

There is a long history of work contributing to the idea that J0737$-$3039B formed via electron capture supernova. \cite{plp+04} first suggested that the critical stellar mass required to form a NS (\unit[10$-$12]{\Msun} for solitary stars) should be significantly lower for tight, interacting binary systems (\unit[6$-$8]{\Msun}). Based on early scintillation velocity measurements, contraints were placed on the progenitor mass of J0737$-$3039B and kick velocity due to its supernova \citep{wk+04, ps+05, wkh+05, wkf+06}. Precise transverse velocity measurements from an extended timing campaign provided the necessary information to claim an unusually low progenitor mass for J0737$-$3039B and corresponding low supernova kick velocity \citep{std+06,ps+06}.

We draw attention to the double pulsar system here to illustrate the detailed process that is required to make claims about DNS formation scenarios. As a result of the mass constraints presented in this paper, \jname\ appears to be in a DNS system. Its unique spin and orbital parameters challenge models that describe DNS formation.

In \S\ref{sec:tim}, we describe the GBT \unit[350]{MHz} Drift Scan survey and the Pulsar Search Collaboratory, as well as the follow-up timing campaign and the parameters measured for \jname; \S\ref{sec:companion} provides evidence that a NS companion is likely, although radio follow-up observations have not provided any evidence of a pulsar companion. In \S\ref{sec:conc}, we draw conclusions from our findings and outline plans for future work.

\section{Timing Observations \& Analysis}\label{sec:tim}
In May--August of 2007, when the Green Bank Telescope (GBT) was undergoing track replacement, the GBT 350-MHz Drift Scan Pulsar Survey \citep{blr+13,lbr+13} looked for radio pulsars as the sky drifted overhead. Of the \unit[1,491]{hours} of recorded drift scan data, \unit[$\sim$300]{hours} were allocated to the Pulsar Search Collaboratory\footnote{The PSC \citep{rhm+10} aims to interest high-school students in science, technology, engineering and mathematics (STEM) related career paths, focusing especially on engaging women and minority students as well as those from low-income families.} (PSC). The survey and follow-up timing observing campaigns, processing pipeline and the students' first five discoveries are discussed in detail in \cite{rsm+13}. \psr\ is the sixth pulsar discovered by PSC students.

\begin{figure}[t!]
\begin{center}
\includegraphics[scale=0.6,angle=0]{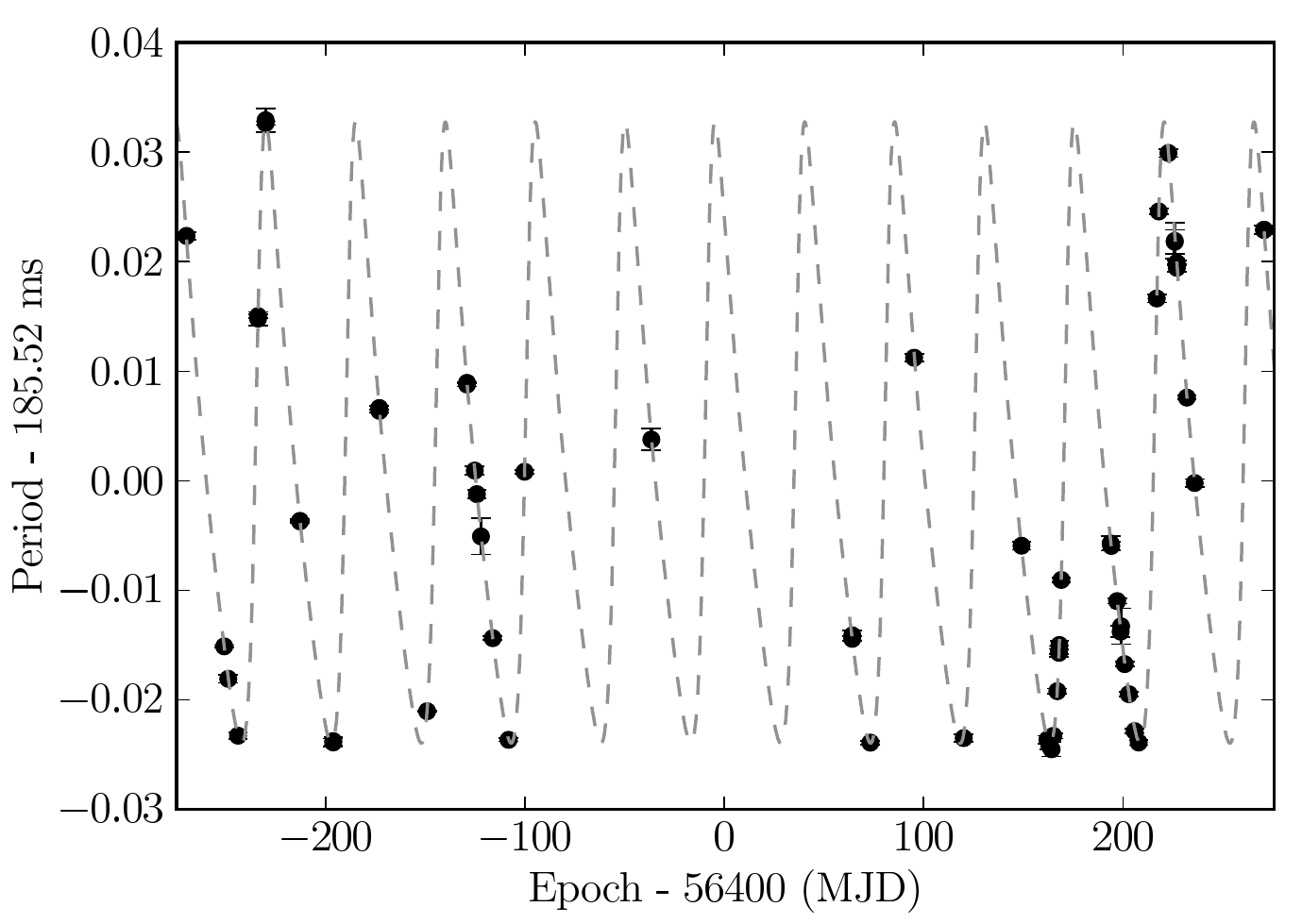}
\caption{Black points show spin period measurements at various observing epochs with error bars, often smaller than the points themselves. The gray dashed line illustrates predicted spin period versus time based on our five Keplerian, orbital parameters. The apparent spin period variation shown here is a Doppler effect due to binary motion and the pulsar's intrinsic spin period is given in Table \ref{tab:par}.}
\label{fig:orbit}
\end{center}
\end{figure}

\begin{deluxetable}{ll}[p]
\tablecaption{\label{tab:par}Timing parameters for \psr}
\tablewidth{220pt}
\tablecolumns{2}      
\tablehead{\colhead{Measured Parameters~~~~~~~~~~~~~~~~~~~~~} & \colhead{Value}}   
\startdata   

Right Ascension (J2000)\dotfill & 19:30:29.7156(7) \\
Declination (J2000)\dotfill & -18:51:46.27(6) \\
Spin Period (s)\dotfill & 0.18552016047926(8) \\
Period Derivative (s/s)\dotfill & 1.8001(6)$\times 10^{-17}$ \\
Dispersion Measure (\dmu)\dotfill & 42.8526(4) \\
Reference Epoch (MJD)\dotfill & 56513 \\
Span of Timing Data (MJD)\dotfill & 56121$-$56904 \\
Number of TOAs\dotfill & 75 \\
RMS Residual (\us)\dotfill & 29 \\
$\chi^2_{\rm red}$\dotfill & 1.05 \\
\smallskip \\
Binary Parameters & \\
\hline \\
Orbital Period (days)\dotfill & 45.0600007(5) \\
Projected Semi-major Axis (lt-s)\dotfill & 86.890277(7) \\
Epoch of Periastron (MJD)\dotfill & 56526.642330(3) \\
Longitude of Periastron (deg)\dotfill & 292.07706(2) \\
Orbital Eccentricity\dotfill & 0.39886340(17) \\
Advance of Periastron (deg/yr)\dotfill & 0.00078(4) \\
\smallskip \\
Derived Parameters & \\
\hline \\
Surface Magnetic Field (10$^{10}$ Gauss)\dotfill & 6.0 \\
Spin-down Luminosity (10$^{32}$ erg/s)\dotfill & 1.1 \\
Characteristic Age (Myr)\dotfill & 163 \\
Mass Function (\Msun)\dotfill & 0.34690765(8) \\
Minimum Companion Mass (\Msun)\dotfill & 1.30\footnote{Minimum companion mass listed here is based on constraints provided by $M_{\rm tot}$ (see \S\ref{sec:mcs}).} \\
Combined Mass (\Msun)\dotfill & 2.59(4) \\
Mean $S_{820}$ (mJy)\dotfill &0.7 \\
\enddata
\tablecomments{Uncertainties in the last significant digit(s) are quoted in parentheses and represent $1\sigma$ errors on measured parameters. Since the flux density ($S_{820}$) quoted here is based on a single-epoch measurement, the uncertainty may be 10$-$20\%.}
\label{(text)}
\end{deluxetable}

\begin{figure*}
\begin{center}
\includegraphics[scale=0.5,angle=0]{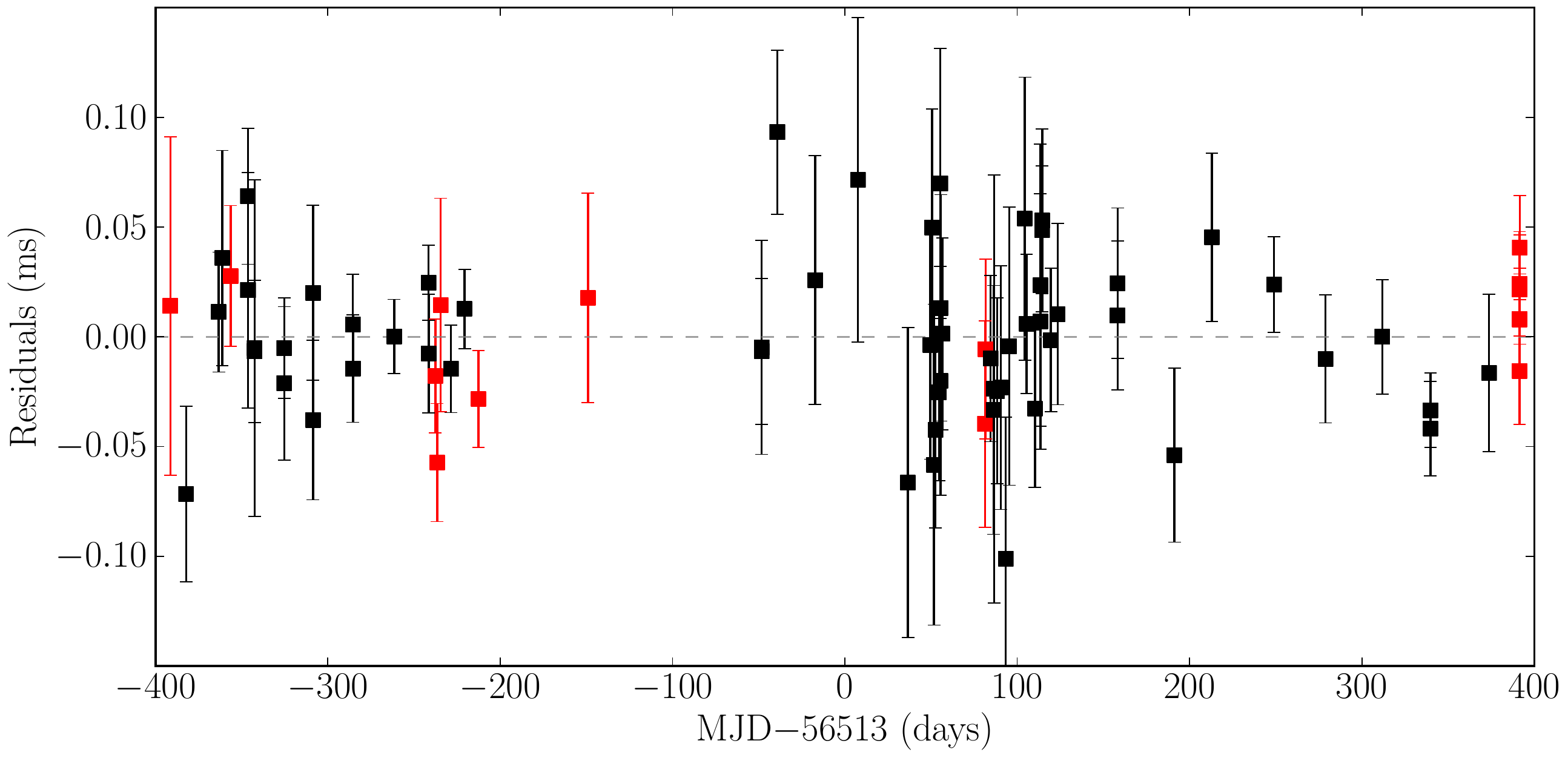}
\caption{Timing residuals for \jname\ plotted here correspond to fit parameters listed in Table \ref{tab:par}. Red and black points represent \unit[350]{MHz} and \unit[820]{MHz} observations respectively. The group of \unit[350]{MHz} TOAs at MJD 56904 come from a \unit[3.5]{hr} observation at superior conjunction.}
\label{fig:resid}
\end{center}
\end{figure*}

After \jname\ was flagged by PSC students in late July of 2012, we confirmed the candidate with a long scan at \unit[350]{MHz} with the GBT, at which point we also verified the binary nature of the source, given the significantly different measured spin period than that reported on the discovery plot. Soon after confirmation, we performed ``gridding" observations, tiling the \unit[350]{MHz} beam with seven \unit[820]{MHz} beam positions, since the higher-frequency receiver has a smaller angular beam size on the sky (\unit[$\sim$0.25]{deg} compared to \unit[$\sim$0.5]{deg}). As described by \cite{mhl+02}, gridding reduces the uncertainty on the pulsar's position and makes it easier to eventually achieve a phase-connected timing solution. At the GBT, the \unit[350]{MHz} prime-focus receiver is usually only mounted for several days each month, while the \unit[820]{MHz} receiver is up for the rest of the month, so gridding also provided more flexibility in our follow-up timing campaign.

We first conducted high-cadence, then monthly timing observations once we had an orbital solution. We observed \jname\ at \unit[350]{MHz} and (primarily) \unit[820]{MHz} center frequencies, with \unit[100]{MHz} and \unit[200]{MHz} of bandwidth respectively. For all observations, we used the Green Bank Ultimate Pulsar Processing Instrument (GUPPI; \cite{drd+08}) with \unit[81.92]{$\mu$s} resolution time and 2048 frequency channels.

During each session, we observed \jname\ for \unit[$\sim$15]{mins}, manually excised RFI with {\tt psrzap} (part of the {\sc PSRCHIVE}\footnote{http://psrchive.sourceforge.net} software package, \cite{hvm+04}) and then summed the signal across the entire bandwidth. We summed across the time domain to generate one mean pulse profile per session and compared it with a synthetic standard profile to compute a time of arrival (TOA) using the {\sc PSRCHIVE} routine {\tt pat}. Standard profiles, one for each observing frequency, were created by fitting Gaussian components to a high signal-to-noise profile.

We measured a spin period at each observing epoch then, using methods from \cite{bn+08}, found an orbital solution for \jname; period measurements and those predicted by our orbital solution are shown in Figure \ref{fig:orbit}. We further refined orbital parameters in {\sc Tempo2}\footnote{http://tempo2.sourceforge.net} \citep{t2} to achieve a full, phase-connected solution, shown in Table \ref{tab:par}. Our timing solution, in Barycentric Coordinate Time (TCB), uses the DE405 Solar System ephemeris and TT(BIPM) clock corrections. Subtracting TOAs from modeled arrival times determined by our phase-connected timing solution yields timing residuals shown in Figure \ref{fig:resid}. Since the reduced chi-squared statistic, $\chi^2_{\rm red}\sim1$ for our timing residuals, we do not use a multiplicative ``error factor'' (EFAC), so uncertainties given in Table \ref{tab:par} are identical to those reported by {\sc Tempo2}.
  
On September 4, 2014, we observed \jname\ at superior conjunction (orbital phase, $\phi\sim0.373$) for \unit[3.5]{hrs} with our normal \unit[350]{MHz} setup described above. We obtained several TOAs from this epoch \--- displayed in Figure \ref{fig:resid} and included in the timing solution reported in Table \ref{tab:par}. Since the post-fit residuals did not show any sign of a Shapiro delay signature, we do not fit for Shapiro delay range ($r$) and shape ($s$) parameters here.

\begin{figure}
\begin{center}
\includegraphics[scale=0.6,angle=0]{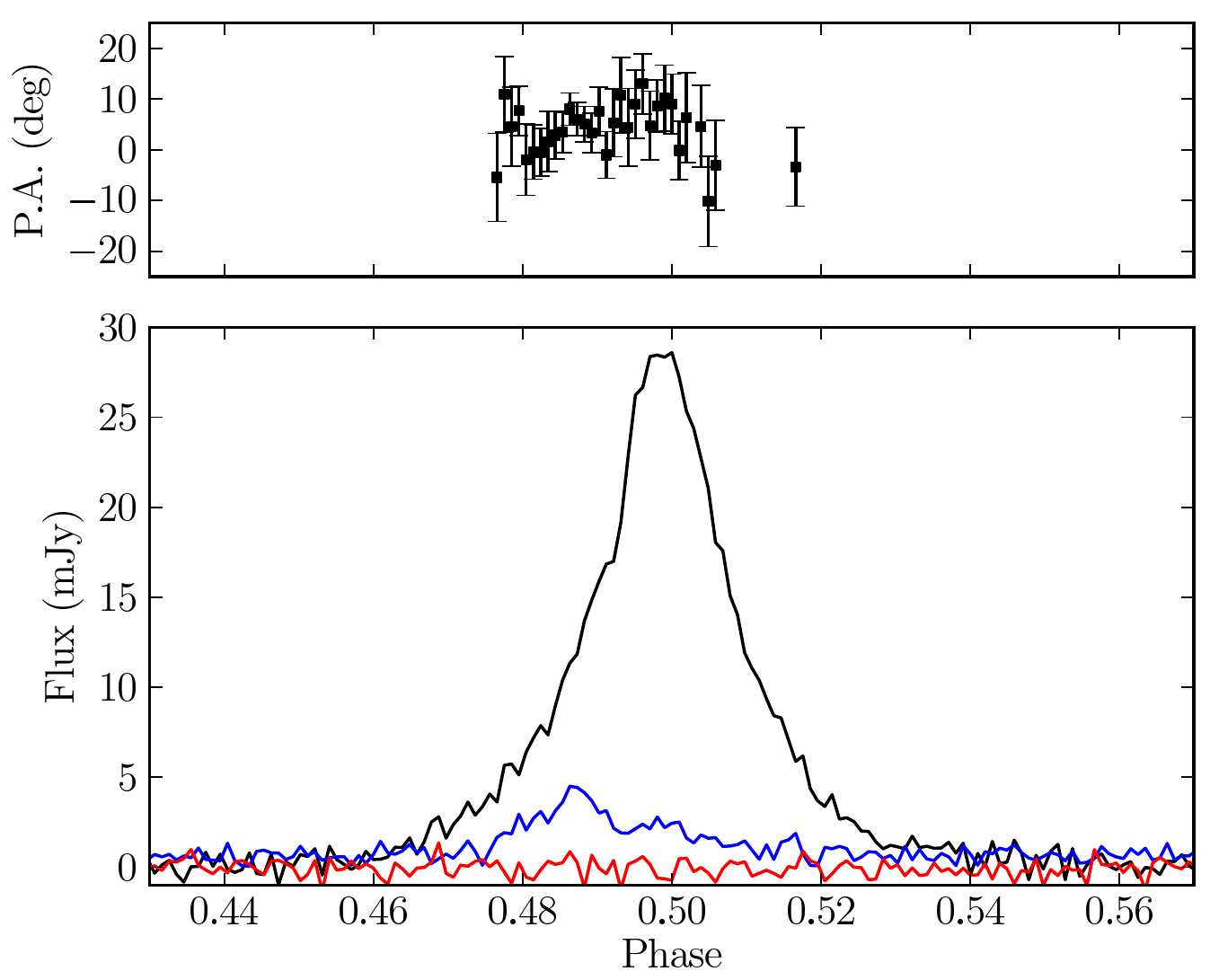}
\caption{{\it Bottom panel:} A coherently dedispersed, flux- and polarization-calibrated pulse profile for \jname\ obtained from a \unit[1]{hr} observation at \unit[820]{MHz} with \unit[200]{MHz} bandwidth and 1024 profile bins. The black line represents total intensity, while linear and circular polarization are shown in blue and red respectively. {\it Top panel:} The position angle (P.A.) swing due to slight linear polarization in the leading edge of the pulse profile.}
\label{fig:profile}
\end{center}
\end{figure}

A flux- and polarization-calibrated, 820-MHz pulse profile for \jname\ is shown in Figure \ref{fig:profile}. About 15\% of \jname's emission is linearly polarized, but due to the relatively flat position angle curve, we were not able to fit this curve using the rotating vector model \citep{rc+69}. There is no trace of circularly-polarized emission.

\section{Nature of the Companion}\label{sec:companion}
Given optical images of the sky surrounding \jname\ and mass constraints based on our timing solution (Table \ref{tab:par}), the pulsar's companion is most likely another neutron star.

\subsection{Optical Follow-up}
Assuming the companion is a main sequence star, we can estimate its apparent bolometric magnitude and apparent SDSS g magnitude.  First, with the mass-luminosity relation for an appropriate mass range $L_{\rm c,min}/{\rm L_\odot}=(m_{\rm c,min}/{\rm M_\odot})^{3.5}$ \citep{allen+73}, where $m_{\rm c,min}$ is the minimum companion mass, we find \unit[$L_{\rm c,min}=2.5$]{\Lsun}, which would correspond to an F5V spectral type.  Next, using a DM-estimate distance, \unit[$d_{\rm DM}\sim1.5$]{kpc} \citep{ne2001} we estimate an apparent bolometric magnitude of $m_{\rm bol,c}=14.7$. To convert this to an apparent g magnitude, we (i) applied a bolometric correction (${\rm BC}\sim-0.09$) to convert $m_{\rm bol,c}$ to an apparent V magnitude \citep{bcp+98}, (ii) transformed this V magnitude to a SDSS g magnitude \citep{jsr+05}, and (iii) included the effects of extinction (\unit[$A_{\rm g}\sim0.33$]{mag}) using the estimates of \cite{sf+11}.  If the companion is a main sequence star, we find that it should have an apparent magnitude of $g_{\rm c}=15.27$ or brighter.

\begin{figure}[t!]
\begin{center}
\includegraphics[scale=0.8,angle=0]{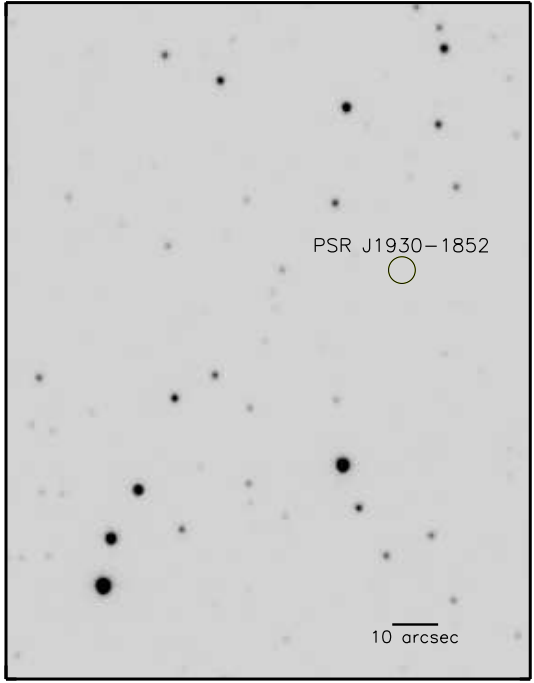}
\caption{This 1.9$'\times$2.5$'$ field of view surrounding \jname\ is the result of stacking 20 \unit[300]{s} images taken with a SDSS $g$ filter and the CTIO \unit[0.9]{m} telescope. No objects were detected within a five-arcsecond radius around \jname\ --- well beyond the uncertainties in measured position.}
\label{fig:opt}
\end{center}
\end{figure}

We observed the field around \jname\ on the night of 9 May 2014 using the CTIO \unit[0.9]{m} telescope, which was accessed through the SMARTS \citep{SMARTS} Consortium.  We obtained 20 images of the field through a SDSS g filter (``CTIO 4770/1006'') over the course of two hours; each individual exposure had an exposure time of 300 seconds, giving us a total integration time of 6000 seconds.  To minimize processing time, we read out only a 291$\times$375 pixel subsection of the full 2048$\times$2048 pixel CCD, which, with a \unit[0.401]{$''$/pixel} plate scale, gave us a 1.9$'\times$2.5$'$ field of view (see Figure \ref{fig:opt}).

All frames were flat-fielded and bias-subtracted using standard routines in IRAF\footnote{IRAF is distributed by the National Optical Astronomy Observatories, which are operated by the Association of Universities for Research in Astronomy, Inc., under cooperative agreement with the National Science Foundation.} \citep{IRAF}, and the 20 reduced object frames were averaged together into a master frame with the {\tt imcombine} task.  We used {\it astrometry.net} to perform an astrometric calibration of the master frame to a precision of better than 0.1$''$. To place an upper limit on the magnitude of any optical signatures, we first determined the aperture diameter that maximized the signal-to-noise (S/N) ratio in the photometry of nearby stars (2.4$''$, or about two times the seeing); next, we calculated the number of counts in the area of sky centered on \jname\ and determined the level of noise at this position. Setting a 5$\sigma$ detection threshold requirement for faint stars in the vicinity, we find a limiting magnitude of $g = 20.5$ for a possible optical counterpart of \jname.

Since the estimated g magnitude for a MS star companion $g_{\rm c}=15.27$ is $\sim$125 times brighter than the limiting magnitude achieved here, we rule out a MS companion. The uncertainty in $d_{\rm DM}$ ($\sim25\%$) does not alter this conclusion.

\subsection{Mass Constraints}\label{sec:mcs}
The mass function expresses the mass of the pulsar ($m_{\rm p}$) and that of the companion ($m_{\rm c}$) in terms of Keplerian orbital parameters $a_{\rm p}\sin i$ (projected semi-major axis) and $P_{\rm b}$ (orbital period). Using those measured parameters and setting the inclination angle $i=90^{\circ}$, we place a lower limit on $m_{\rm c}$ for any given $m_{\rm p}$.

The measurement of the relativistic advance of periastron ($\dot{\omega}$) also provides the system's total mass, \unit[$M_{\rm tot}=2.59(4)$]{\Msun}. Taking $i=90^\circ$, we use the mass function and double the uncertainties on $M_{\rm tot}$ to place $2\sigma$ lower/upper limits on the mass of the companion (\unit[$m_c\geq1.30$]{\Msun}) and the mass of the pulsar (\unit[$m_p\leq1.32$]{\Msun}) respectively.

Based on the companion mass lower limit \unit[$m_{\rm c, LL}=1.30$]{\Msun}, the orbital eccentricity $e\sim0.4$, the spin period and period derivative of \jname, a NS is the most likely companion.

\subsection{Radio Follow-up}
We dedispersed our \unit[3.5]{hr} superior conjunction observation using the DM measured for \jname\ (\unit[${\rm DM}=42.85$]{pc~cm$^{-3}$}), took a discrete Fourier Transform of the resulting timeseries and performed an acceleration search for a possible pulsar companion. To do so, we used the {\tt accelsearch} routine from the {\sc Presto}\footnote{https://github.com/scottransom/presto} suite of pulsar search software \citep{ransom+01}, examining the frequency domain with a matched filter template up to 20 Fourier bins wide. To maximize our sensitivity to pulsars with narrow profiles, we summed up to 16 harmonics for candidate signals. This procedure returned over 800 high-significance (S/N > 9) candidates, from which we removed obvious RFI and those that were harmonically related to \jname\ or each other.\

We folded and visually inspected the remaining candidates, but did not find any evidence of a pulsar counterpart. Assuming the harmonic summing was close to ideal, we were sensitive to ${\rm S/N}>9$ signals in the time domain, which corresponds to a \unit[350]{MHz} flux limit of  \unit[$\sim$30]{$\mu$Jy}, given our observing set-up. These results suggest that a possible pulsar companion is either too weak to be detected or is not beaming along our line of sight. Given the sensitivity limit reached, the latter explanation is more likely.

\begin{deluxetable*}{lccccccccc}[h!]
\tablewidth{500pt}
\tablecaption{\label{tab:dns}Known Recycled Pulsars in DNS Systems}
\tablecolumns{9}
\tablehead{\colhead{Pulsar} & \colhead{$P_{\rm spin}$} & \colhead{$\dot{P}$} & \colhead{$e$} & \colhead{$P_{\rm orb}$} & \colhead{$m_{\rm p}$} & \colhead{$m_{\rm c}$} & \colhead{$M_{\rm tot}$} & \colhead{Recent} \\
 & (ms) & (\unit[$10^{-18}$]{s/s}) & & (days) & (\Msun) & (\Msun) & (\Msun) & References }
\startdata
J0737$-$3039A & 22.7 & 1.8 & 0.09 & 0.10 & 1.34 & 1.25 & 2.59 & \cite{ksm+06} \\
J1756$-$2251 & 28.5 & 1.0 & 0.18 & 0.32 & 1.34 & 1.23 & 2.57 & \cite{fsk+14} \\
B1913+16 & 59.0 & 8.6 & 0.62 & 0.32 & 1.44 & 1.39 & 2.83 & \cite{wnt+10} \\ 
B1534+12 & 37.9 & 2.4 & 0.27 & 0.42 & 1.33 & 1.35 & 2.68 & \cite{fst+14} \\
J1829+2456 & 41.0 & 0.05 & 0.14 & 1.18 & $<1.34$ & $>1.26$ & 2.53 & \cite{clm+04,clm+05} \\
J0453+1559 & 45.8 & 0.19 & 0.11 & 4.07 & 1.54 & 1.19 & 2.73 & Martinez et al. (in prep.) \\
J1518+4904 & 40.9 & 0.03 & 0.25 & 8.63 & $<1.17$ & $>1.55$ & 2.72 & \cite{jsk+08} \\ 
J1753$-$2240 & 95.1 & 1.0 & 0.30 & 13.6 & $-$ & $-$ & $-$ & \cite{kkl+09} \\
J1811$-$1736 & 104 & 0.9 & 0.83 & 18.8 & $<1.64$ & $>0.93$ & 2.57 & \cite{cor+07} \\
J1930$-$1852 & 186 & 18 & 0.40 & 45.1 & $<1.32$ & $>1.30$ & 2.59 & ---
\enddata
\tablecomments{A comparison between \jname\ and all other known primary, partially-recycled DNS pulsars, sorted by $P_{\rm orb}$. PSRs J1906+0746 and J0737$-$3039B were omitted because neither underwent recycling \citep{lsf+06,ksm+06}. PSR B2127+11C \citep{jcj+06} was also omitted because it was formed in a globular cluster, indicating a different evolutionary history.}
\end{deluxetable*}

\begin{figure*}
\begin{center}
\includegraphics[scale=1.0,angle=0]{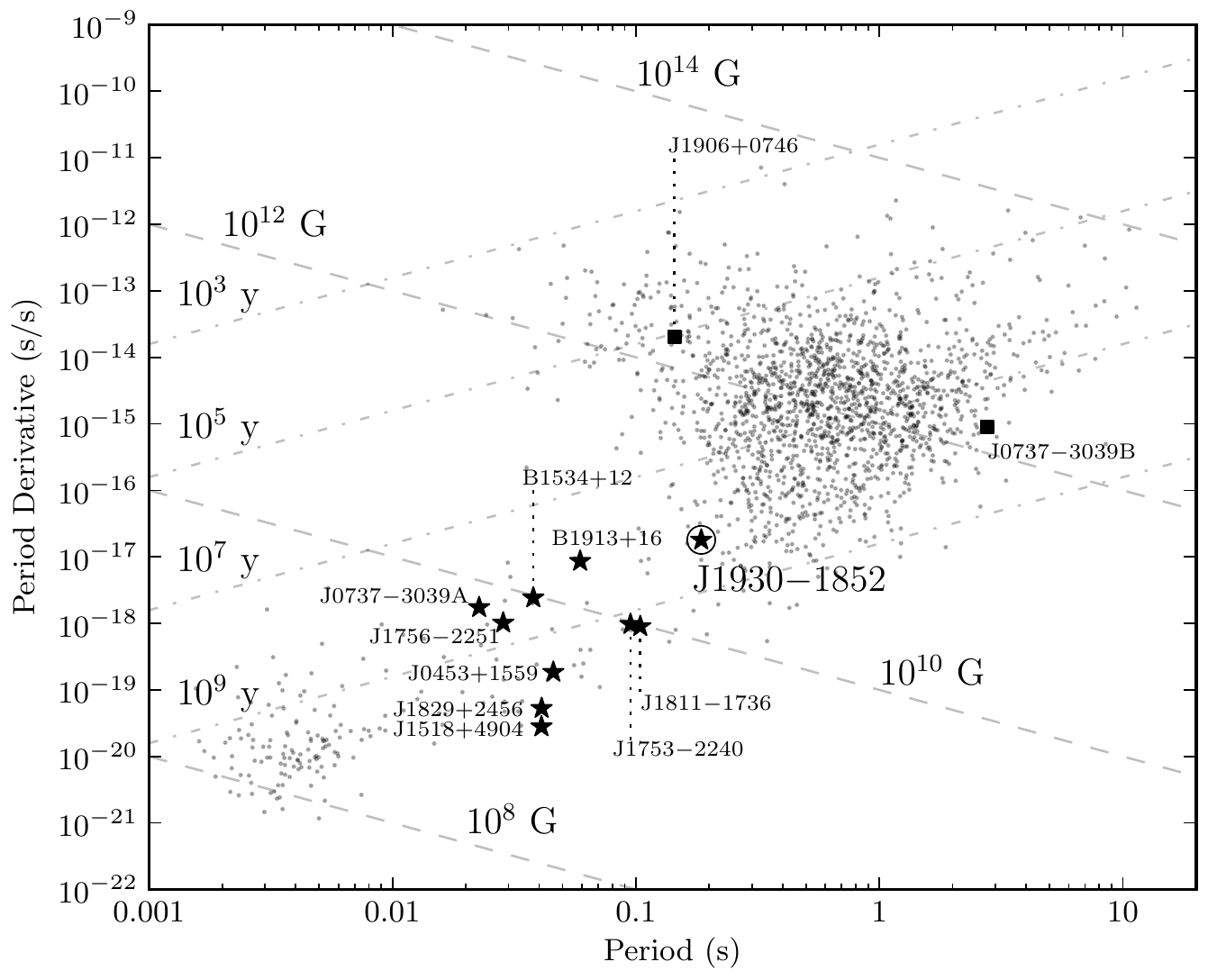}
\caption{A $P$--$\dot{P}$ diagram showing all pulsars in DNS systems (stars/squares) and all other known pulsars (dots). Measured $P$ and $\dot{P}$ come from the ATNF Pulsar Catalog \citep{psrcat} and lines of characteristic age and surface magnetic field are shown with dot-dash and dashed lines, respectively. Recycled DNS pulsars (stars) appear between the normal and millisecond pulsar populations and are listed in Table \ref{tab:dns}. Despite its significantly longer spin period, \jname\ clearly belongs in the population of recycled DNS pulsars, unlike J1906+0746 and J0737$-$3039B (squares) \--- neither of which have undergone recycling.}
\label{fig:ppdot}
\end{center}
\end{figure*}

\begin{figure}
\begin{center}
\includegraphics[scale=0.6,angle=0]{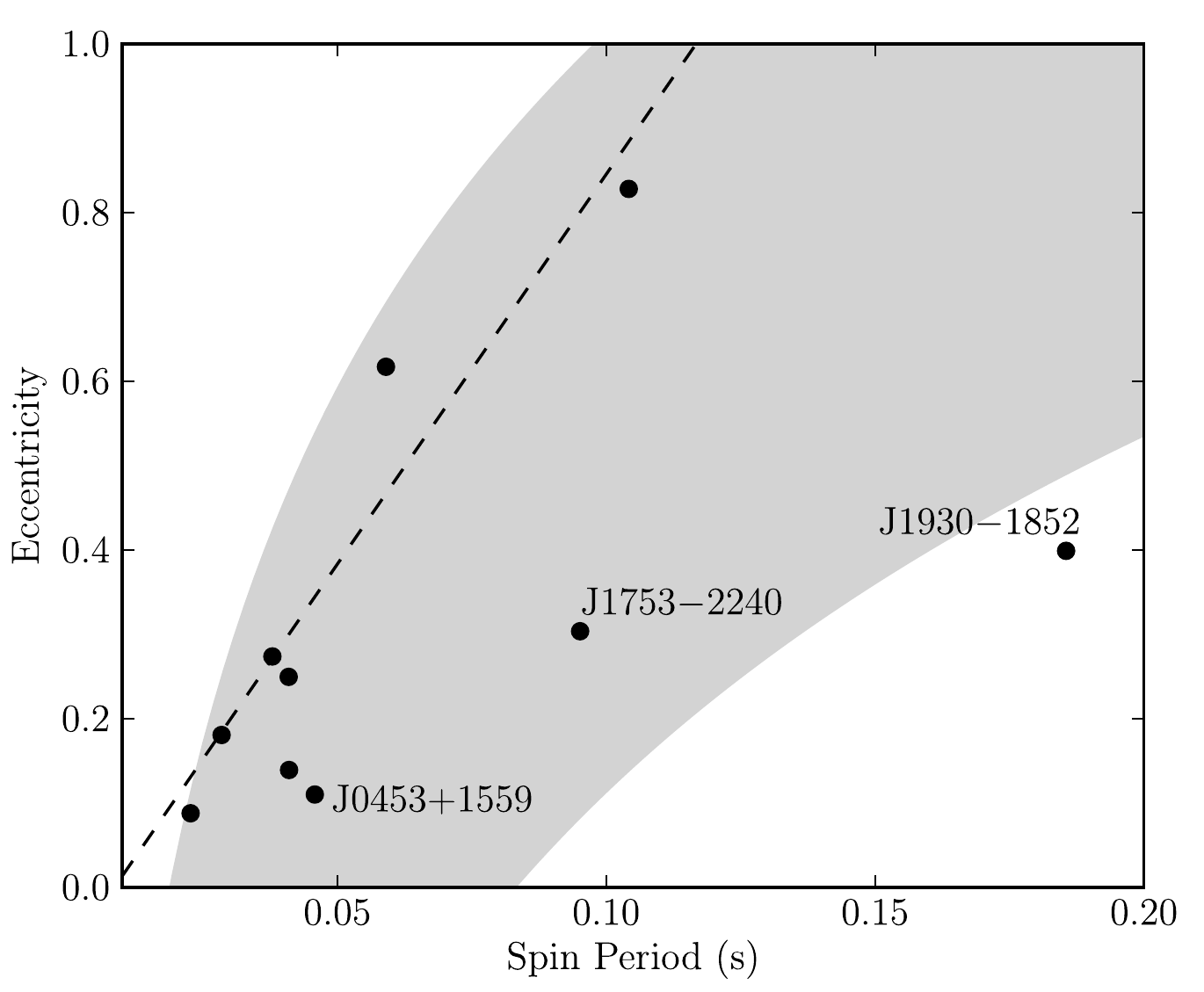}
\caption{The dashed line shows the spin period\--eccentricity relation first noted by \cite{mlc+05} and \cite{fkl+05}, which is simply a fit using the first seven known primary, partially-recycled DNS pulsars (unmarked points). The shaded region shows the approximate spread in properties of simulated DNS systems according to \cite{dpp+05}, assuming a small Maxwellian kick velocity dispersion induced by the second supernova (\unit[$\sigma_2=20$]{km~s$^{-1}$}).\footnote{Although we were unable to obtain actual population synthesis results from Figure 3 in that paper, we estimated the spread in spin period and eccentricity for simulated DNS systems by eye.} PSRs J1753$-$2240 and \jname\ have been labeled because they both fall off the dashed line, but have spin periods and eccentricities that are still roughly consistent with the \cite{dpp+05} synthetic population. PSR J0453+1559, a recently discovered DNS with a large mass asymmetry,  described by Martinez et al. (in prep.), is consistent with the shaded region as well as the distribution of previously-discovered DNS pulsars shown here.}
\label{fig:pe}
\end{center}
\end{figure}

\section{Summary \& Conclusions}\label{sec:conc}
Analysis presented in \S\ref{sec:companion} implies that \jname's most likely counterpart is another NS. We compare \jname's parameters with those of other pulsars in known DNS systems that have similar evolutionary scenarios in Table \ref{tab:dns}. Given \jname's moderately short spin period and period derivative well below those measured for otherwise similar unrecycled pulsars, \jname\ is partially recycled and therefore, most likely formed before its companion. On a $P-\dot{P}$ diagram (see Figure \ref{fig:ppdot}), \jname\ falls in the same region as other recycled DNS systems. Meanwhile, it has a longer spin period (\unit[$P_{\rm spin}\sim185$]{ms}) and higher rate of spin-down (\unit[$\dot{P}\sim2\times10^{-17}$]{s/s}) than any other first-born, recycled DNS pulsar and an orbital period (\unit[$P_{\rm b}\sim45$]{days}) longer than any other DNS system. These together may imply a shorter than average and/or inefficient mass transfer phase before the companion went supernova. Since accretion is thought to be the source of significant mass gain for recycled NSs \citep{zwz+11}, a short and/or inefficient accretion period is consistent with the relatively low upper limit we place on the mass of \jname, \unit[$m_{\rm p}<1.32$]{\Msun}. The upper mass limit for \jname\ is lower than those of all other recycled DNS pulsars, except J1518+4904. \\

\subsection{Spin Period\---Eccentricity Relationship}\label{sub:pe}
The $P_{\rm spin}-e$ relationship was first noted in \cite{mlc+05} and \cite{fkl+05}, after the discoveries of J1756$-$2251 and J1829+2456; \cite{fkl+05} performed a linear fit in spin period/eccentricity space (see Figure \ref{fig:pe}), using the first seven known primary, partially recycled DNS pulsars and found a Pearson correlation coefficient $r=0.97$. Monte Carlo simulations showed that such high $r$ values would only be expected to occur by chance 0.1\% of the time.

In the absence of a supernova kick, a positive correlation between eccentricity and spin period is expected. Without an appreciable kick, the DNS system's eccentricity following the second supernova explosion is directly proportional to the mass lost in the process \citep{wak+08}. A lower-mass secondary will evolve more slowly, prolonging the mass transfer phase and allowing the primary more time to accrete material and spin up. The low-mass secondary loses relatively little mass during its supernova explosion, so the resulting DNS system (if it remains bound) has low eccentricity and a primary NS with a short spin period. Conversely, a high-mass secondary evolves more quickly. A short mass transfer phase before the secondary's supernova leaves the primary with a long spin period; greater mass loss during the supernova results in a DNS system with high eccentricity.

Using population synthesis simulations of DNS formation, \cite{dpp+05} show that this expected $P_{\rm spin}-e$ relationship persists for small supernova kick velocities. The authors assume standard and double-core formation scenarios and that most of the matter accreted by the primary NS comes from a helium star companion during the He-star\---NS binary mass transfer phase. Finally, they assume that the accretion rate is limited by the Eddington rate for helium accretion and invoke a model to account for accretion-induced spin-up of the primary NS. By applying a small Maxwellian kick velocity distribution with dispersion \unit[$\sigma_2=20$]{km~s$^{-1}$} for the secondary supernova explosion, the distribution of simulated DNS systems exhibits a slope similar to that of the empirical $P_{\rm spin}-e$ relationship first noted by \cite{mlc+05} and \cite{fkl+05}. The $P_{\rm spin}-e$ relationship is maintained in the simulated population for \unit[$\sigma<50$]{km~s$^{-1}$}, regardless of the kick velocity imparted by the primary supernova.

We plot a representation of the simulated DNS population from \cite{dpp+05} as a shaded region in Figure \ref{fig:pe}. \cite{kkl+09} used a similar representation to argue that J1753$-$2240 is perhaps more representative of the DNS population than previously discovered DNS pulsars, which all lie closer to the more sparsely-populated, low-$P_{\rm spin}$ edge of the shaded region. Although \jname\ appears to be approximately consistent with the \cite{dpp+05} simulations, its location in a sparsely-populated portion of the proposed $P_{\rm spin}-e$ distribution draws the model's assumptions and conclusions into question.

In their paper, Dewi et al. note that the $P_{\rm spin}-e$ relationship is destroyed (randomized) in a simulated population of DNS systems with high supernova kick velocities (\unit[$\sigma>50$]{km~s$^{-1}$}). There is good evidence \citep[e.g.][]{wwk+10} that both B1913+16 and B1534+12 require high supernova kicks, possibly \unit[$\gg$150]{km~s$^{-1}$}, but these DNS systems remain completely consistent with the simulated distribution of systems with low kick velocites shown in Figure \ref{fig:pe}. Therefore, there is currently no known DNS system that supports the claim from \cite{dpp+05} that high supernova kick velocity systems do not follow a $P_{\rm spin}-e$ relationship. However, \jname\ (like J1753$-$2240) provides further evidence that any relationship between spin period and eccentricity is probably much broader than was originally thought.

\subsection{Orbital Period-Eccentricity Relationship}\label{sub:pbe}
For short orbital period binary systems, one would expect a prolonged or more efficient recycling process than for long orbital period systems. Since the amount of recycling that occurs is inversely related to both orbital period and spin period, these quantities are expected to track one another (i.e. long orbital period DNS binaries should have long spin periods). We see this relationship in the known DNS population and Table \ref{tab:dns} illustrates it nicely. Building on arguments in \S\ref{sub:pe}, we expect DNS systems with longer orbital periods to also have larger eccentricities.

\cite{afk+15} simulate DNS system formation, considering three dominant evolutionary channels for the primary NS's companion:
\begin{inparaenum}[(i)]
\item wide-orbit common envelope, followed by iron core-collapse supernova,
\item tight-orbit common envelope, followed by a second round of mass transfer, then iron core-collapse supernova, and 
\item similar to channel ii, but a lower-mass secondary He core forms a NS through electron capture supernova.
\end{inparaenum}
Using currently understood evolutionary histories for PSRs J0737$-$3039A/B, B1534+12 and B1913+16, the authors constrain DNS population models and binary parameters. For each synthetic population representing a given model, they compare confidence intervals that result from simulations to the actual DNS population in the $P_{\rm b}-e$ plane. Confidence intervals for all evolutionary channels combined follow a roughly positive slope in the $P_{\rm b}-e$ plane, as one would expect. High-confidence regions shift depending on the chosen set of model parameters and evolutionary channel. For example, higher supernova kick velocities tend to produce systems with higher eccentricities; decreasing the efficiency of the common envelope ejection mechanism produces systems with wider orbits. Of the three evolutionary channels, the first typically has the broadest distribution in orbital period. In most cases (i.e. for most chosen sets of model parameters), the known DNS population falls well within 3$\sigma$ confidence intervals considering all evolutionary channels (i, ii and iii) combined. Comparing \jname's parameters to the confidence intervals shown for the reference model in \cite{afk+15} (Figure 1 in that paper), the combination of orbital period and eccentricity is inconsistent with simulated parameters of systems from the three dominant evolutionary channels at the $4-5\sigma$ level. Many models in \cite{afk+15} do not produce systems like \jname, which suggests that it may be useful for constraining theory.

\subsection{Future Work}\label{sub:fw}
\psr's parameters set it apart from previously-studied primary DNS pulsars: the widest orbit (longest orbital period), the longest spin period, the largest spin-down rate. We hope to determine the pulsar and companion masses independently and accurately measure proper motion to infer (i) the progenitor mass of the second-born NS and (ii) the magnitude of the supernova kick it received at birth. Then, following work by \cite{wwk+10}, we can make meaningful predictions about \jname's true evolutionary history. 

So far, we detect no Shapiro delay signature in our timing residuals (see Figure \ref{fig:resid}), despite having good orbital coverage and a \unit[3.5]{hr} superior conjunction observation. This non-detection could be a result of a combination of factors including: inadequate timing precision, low companion mass and/or small inclination angle. If the non-detection is related to timing precision, continued timing observations may result in a measurement of the delay, and hence an additional constraint on the companion mass.

Upcoming VLBA observations will provide high-precision position, proper motion and parallax measurements, which will help isolate spin, orbital and post-Keplerian parameters in our timing residual fits. Also, because \jname\ is relatively nearby (\unit[$d_{\rm DM}\sim1.5$]{kpc}), we will be able to resolve the orbit with VLBA imaging; at the estimated distance, the major axis of \jname's orbit spans \unit[$\sim$250]{$\mu$as} and the VLBA routinely obtains \unit[50]{$\mu$as} precision or better with an in-beam calibrator available \citep{dab+12}. Resolving the orbit will allow us to place constraints on the system's inclination angle and in turn, the masses of \jname\ and its companion. These measurements are critical for eventually understanding the formation scenario for \jname\ and other wide-orbit DNS systems.

\section*{Acknowledgements}
We thank Robert Ferdman, Ingrid Stairs and Jeffrey Andrews for useful discussions. The PSC was funded through NSF ITEST award number 0737641. J.K.S., M.A.M., and D.R.L. are supported through NSF PIRE award number 0968296. T.H., E.F., J.A.M, and B.N.B. acknowledge High Point University, which provided the financial support necessary to conduct the photometric observations on-site at the CTIO 0.9-m telescope. The National Radio Astronomy Observatory is a facility of the National Science Foundation operated under cooperative agreement by Associated Universities, Inc.

\bibliography{1930}
\bibliographystyle{apj}

\end{document}